# PathCo-LatticE: Pathology-Constrained Lattice-Of-Experts Framework for Fully-supervised Few-Shot Cardiac MRI Segmentation

Mohamed Elbayumi[1,2], and Mohammed S.M. Elbaz[1,2*]
Departments of [1]Radiology and [2]Biomedical Engineering, Northwestern University, Chicago, IL, USA
*corresponding author email: mohammed.elbaz@northwestern.edu

*Abstract*— **Few-shot learning (FSL) mitigates data scarcity in cardiac MRI segmentation but typically relies on semi-supervised techniques sensitive to domain shifts and validation bias, restricting zero-shot generalizability. We propose PathCo-LatticE, a fully supervised FSL framework that replaces unlabeled data with pathology-guided synthetic supervision. First, our Virtual Patient Engine models continuous latent disease trajectories from sparse clinical anchors, using generative modeling to synthesize physiologically plausible, fully labeled 3D cohorts. Second, Self-Reinforcing Interleaved Validation (SIV) provides a leakage-free protocol that evaluates models online with progressively challenging synthetic samples, eliminating the need for real validation data. Finally, a dynamic Lattice-of-Experts (LoE) organizes specialized networks within a pathology-aware topology and activates the most relevant experts per input, enabling robust zero-shot generalization to unseen data without target-domain fine-tuning.**

**We evaluated PathCo-LatticE in a strict out-of-distribution (OOD) setting, deriving all anchors and severity statistics from a single-source domain (ACDC) and performing zero-shot testing on the multi-center, multi-vendor M&Ms dataset. PathCo-LatticE outperforms four state-of-the-art FSL methods by 4.2–11% Dice starting from only 7 labeled anchors, approaches fully supervised performance (within 1% Dice) with only 19 labeled anchors. The method shows superior harmonization across 4 vendors & generalization to unseen pathologies. [Code will be made available].**

## I. Introduction

Cardiac Magnetic Resonance Imaging (MRI) is the gold standard for non-invasive assessment of cardiac structure and function [1], [2], [3]. However, its clinical utility relies critically on the accurate segmentation of the left ventricle (LV), right ventricle (RV), and myocardium [4]. These segmentations are prerequisites for quantifying ejection fraction, ventricular volumes, and myocardial mass—parameters indispensable for diagnosing cardiomyopathies and guiding therapy. While Deep Learning (DL) has automated this task [5], its deployment is constrained by a "data bottleneck": fully supervised models require dense, pixel-level expert annotations that are scarce and expensive to acquire. Consequently, models trained on limited data often fail to generalize across diverse pathologies, vendors, and imaging protocols encountered in multi-center practice [6].

To mitigate annotation scarcity, Few-Shot Learning (FSL) has emerged as a promising paradigm. Current FSL approaches in medical imaging [7], [8] predominantly adopt semi-supervised frameworks, such as meta-learning [9] or prototypical networks [7] augmented with unlabeled data. These methods infer class representations from a small labeled "support" set combined with a larger unlabeled cohort. While effective in controlled settings, this reliance on semi-supervision introduces critical limitations for real-world deployment. First, these methods assume access to a representative unlabeled cohort from target domain, which restricts zero-shot generalizability and introduces hidden biases toward the specific distribution of the unlabeled pool. Second, randomly sampled support sets rarely capture continuous spectrum of disease progression, resulting in "spectral gaps" where intermediate or extreme pathological states are underrepresented. Third, common practice of cross-validating on the labeled support set risks data leakage and inflates performance estimates in data-scarce regimes [10], [11].

In this work, we argue that the reliance on unstructured unlabeled data is a bottleneck, not a solution. We propose Pathology-Constrained Lattice-of-Experts (PathCo-LatticE), a framework that reformulates few-shot MRI segmentation from a semi-supervised to a fully supervised, data-centric paradigm.

Our key insight is that cardiac disease progression follows continuous, physiologically measurable trajectories e.g., progressive wall thickening in Hypertrophic Cardiomyopathy (HCM) or ventricular remodeling in Dilated Cardiomyopathy (DCM). Rather than relying on arbitrary unlabeled pools, we introduce Pathology-Constrained Data Synthesis. By identifying a small set of "clinical anchors" at key statistical intervals of disease severity, we train generative models to learn the continuous latent transformations between them. We then sample this latent space to synthesize a large, fully labeled cohort of "virtual patients" ($V$) that uniformly spans the disease spectrum. This effectively replaces the unlabeled set $S_{unlabeled}$ with a synthetic, interpretable, and fully labeled set $V$.

This work was supported in part by National Heart, Lung, and Blood Institute (NHLBI) of the National Institutes of Health (NIH) Grant R01HL169780".

Mohammed S. M. Elbaz and Mohamed Elbayumi are with the Department of Radiology and Biomedical Engineering at Northwestern University, Chicago, IL, USA. Mohammed S.M. Elbaz is the corresponding author: mohammed.elbaz@northwestern.edu

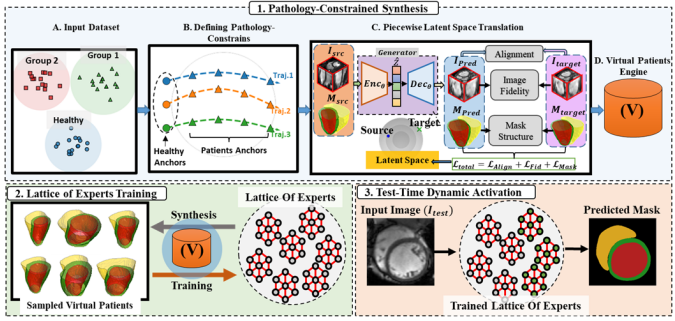

Fig. 1. Overview of PathCo-LatticE. **1) Pathology-Constrained Synthesis:** clinical anchors define key severity levels, and piecewise generators learn 3D image–mask translation, creating a virtual-patient engine. **2) Lattice-of-Experts** is trained on targeted synthetic subsets via clinically guided sampling and self-reinforcement. **3) Test-Time Dynamic Activation:** lattice selects the most confident experts and performs probability voting to produce final output.

To exploit this synthetic cohort, we introduce Self-Reinforcing Interleaved Validation (SIV), a training strategy that validates models on progressively more challenging synthetic samples, eliminating the need for cross-validation on real anchors and preventing leakage. Finally, to handle heterogeneous progression patterns, we construct a Lattice-of-Experts (LoE)—a grid of specialized networks trained at different granularities—coupled with a Test-Time Dynamic Activation mechanism that adaptively selects the optimal expert for each unseen test image. Our main contributions are:

1) *Fully supervised, data-centric FSL*: We reformulate few-shot segmentation by replacing arbitrary unlabeled cohorts with pathology-guided synthetic supervision, addressing class imbalance by filling spectral gaps along the disease continuum.

2) *Pathology-constrained synthesis:* We introduce a *Virtual Patient Engine* that models continuous latent disease trajectories from a few labeled clinical anchors, synthesizing physiologically plausible, fully labeled 3D virtual cohorts as the basis of our fully supervised paradigm.

3) *Leakage-free SIV training:* We propose Self-Reinforcing Interleaved Validation (SIV), which exploits the ordered synthetic progression for online validation, removing the need to reserve real anchors for validation and maximizing the utility of scarce labeled data.

4) *Dynamic Lattice-of-Experts:* We develop a pathology-aware inference architecture that arranges specialized experts in a lattice and dynamically activates the most relevant expert(s) per image/class, enabling zero-shot generalization to unseen vendors and pathologies without target-domain fine-tuning.

5) *State-of-the-art performance:* Using only 7 labeled anchors, PathCo-LatticE outperforms four semi-supervised FSL methods by 4.2–11% Dice on zero-shot out-of-distribution (M&Ms dataset) tests, and with 19 anchors it approaches fully supervised performance on M&Ms (within 1% Dice), while achieving superior multi-vendor harmonization.

## II. RELATED WORK

### A. Generative Models

Early generative models aimed to synthesize realistic images by learning underlying data distributions. Variational Autoencoders (VAEs) [12] produce realistic, smooth outputs, but often lack fine anatomical detail, while Generative Adversarial Networks (GANs) [13] improved visual realism, yet, suffered from mode collapse [14]. In cardiac segmentation, both were mainly used for data augmentation, which often lack structural consistency, limiting their utility for precise segmentation tasks [15]. To improve structural integrity, research shifted to image-to-image translation. Unpaired translators such as CycleGAN [16] enhanced visual realism by employing cycle-consistency loss, yet, remained appearance-driven; they improved domain adaptation but without label supervision. Later, mask-conditioned methods such as XCAT-GAN [17] condition synthesis on segmentation masks to enforce topological correctness. While effective, these methods typically rely on a "two-stage" paradigm—require a separate model for mask manipulation—and fail to model the continuous, non-linear pathology evolution. Recently, Diffusion Models have achieved state-of-the-art fidelity. Recent Contour-guided diffusion [18] generate highly realistic, structurally constrained images. However, these models remain primarily used for augmentation to enhance existing datasets rather than independent sources of fully labeled supervision.

### B. Few-Shot Learning in Cardiac Segmentation

Few-shot learning (FSL) addresses annotation scarcity by generalizing from a limited "support" set. The theoretical foundation was established by Prototypical Networks [19], which perform metric-based classification using class prototypes. In cardiac segmentation, several works extended this idea by incorporating richer prototype formulations [7], [20], or mutual-consistency Parallel efforts in meta-learning [9] adapted optimization-based strategies to segment unseen structures via episodic training. Recent developments have focused on cross-domain generalization: For instance, multi-domain meta-training and meta-transfer approaches [21] have been proposed to handle vendor and disease heterogeneity. However, a critical limitation persists: vast majority of FSL pipelines rely on semi-supervision, requiring a concurrent stream of unlabeled data to align distributions or perform pseudo-labeling [8], [22]. This dependency limits clinical applicability when representative unlabeled cohorts are unavailable and ties performance to hidden biases of the unlabeled pool [23]. Furthermore, standard FSL methods validate on real labeled subjects, introducing data leakage risks in data-scarce regimes [24].

### C. Dynamic Inference and Mixture-of-Experts

Our framework utilizes a "Lattice-of-Experts" to handle heterogeneous pathology. This relates to the concept of Mixture-of-Experts (MoE) [25], where a system decomposes a complex task into sub-tasks handled by specialized networks. In medical imaging, ensemble methods have traditionally aggregated predictions from multiple models to improve robustness. However, standard ensemble is static (applying all models to every input), which is computationally expensive and ignores sample-specific difficulty. More distinct are Test-Time Adaptation (TTA) approaches [26], which fine-tune models on incoming test data. While effective for domain shifts, TTA requires optimization at test-time, which is computationally prohibitive for real-time clinical workflows [27].

*D. Positioning of Our Work*

PathCo-LatticE directly addresses these gaps. Unlike prior generative approaches that serve as stochastic augmentation, it treats generation as pathology-constrained supervision, using continuous disease trajectories to synthesize fully labeled training cohorts and eliminating reliance on unlabeled data. Furthermore, unlike static ensembles or computationally expensive test-time adaptation, our pathology-aware lattice dynamically activates experts in a zero-shot manner based on disease severity, requiring no test-time optimization.

## III. METHODS

*A. Conceptual Framework: Fully-supervised Few-Shot Learning via Pathology-Constrained Synthesis*

We introduce a conceptual reformulation (Fig. 1) that converts the conventional few-shot cardiac MRI segmentation problem from a semi-supervised into a fully supervised setting.

*Traditional Few-Shot Formulation:* In the standard FSL paradigm, the training set is defined as:

$$S_{\text{train}} = S_{\text{few}} \cup S_{\text{unlabeled}} \quad (1)$$

Where $S_{\text{few}}$ is a small, randomly sampled labeled subset and $S_{\text{unlabeled}}$ is a much larger set of unlabeled images. This structure inherently relies on semi-supervised learning or pseudo-labeling and requires cross-validation over a limited number of labeled samples. As a result, the approach is prone to leakage, unstable optimization, and reduced clinical interpretability.

*Our Fully Supervised Reformulation*: replace the semi-supervised structure with a fully supervised $S_{\text{train}}$ defined as:

$$S_{\text{train}} = A \cup V \quad (2)$$

Here, $A$ is a compact set of clinically defined labeled anchors, and $V$ is a large set of fully labeled virtual patients synthesized along disease-specific pathology trajectories. This eliminates the unlabeled component entirely and produces a complete, fully supervised training corpus as follows:

*1) Clinical Anchors and Pathology Progression:*

Each clinical anchor in $A$ is chosen using a disease-specific severity function:

$$f_d: (I, M) \to \gamma \in [0,1] \quad (3)$$

Where $(I, M)$ denotes the 3D labeled image–mask pair and $\gamma$ is a normalized severity value for disease $d$. Unlike generic similarity metrics, $f_d$ is grounded in clinically established biomarkers and reflects the physiological characteristics of the specific pathology (e.g. myocardial mass for HCM, sphericity index for DCM, or ejection fraction for ischemic disease).

The anchor set is therefore formally defined as:

$$A = \{(I_k, M_k, \gamma_k)\}_{k=1}^K, \quad \gamma_k = f_d(I_k, M_k) \quad (4)$$

Where $K$ is intentionally small but the anchors are selected to represent meaningful endpoints or intermediate states along the disease severity spectrum. Anchors form the clinical constraints that govern the subsequent synthesis of virtual patients.

*2) Latent Pathology Trajectory for Virtual Patients*

The core component of our framework is the construction of a continuous latent pathology trajectory:

$$T_d: [0,1] \to (I, M) \quad (5)$$

Which models how both image appearance and anatomical structure evolve as disease severity increases. The trajectory is estimated as a piecewise latent translation between consecutive clinical anchors in $A$. For each anchor $(I_k, M_k)$ with severity $\gamma_k$, the trajectory satisfies:

$$T_d(\gamma_k) \approx (I_k, M_k) \quad \forall k \quad (6)$$

Ensuring that the generative process is tied directly to clinically meaningful endpoints. To populate the pathological continuum between anchors, we sample $T_d$ at a prescribed clinical granularity $\Delta\gamma$. Let:

$$\mathcal{D} = \{p = k\Delta\gamma \mid k \in \mathbb{N}, \; p < 1\} \quad (7)$$

Denote the set of progression states obtained from uniform sampling. The virtual patient set is then defined as:

$$V = \{T_d(p) \mid p \in \mathcal{D}, \; p \notin \{\gamma_k\}_{k=1}^K\} \quad (8)$$

i.e. all synthesized severity states excluding the true anchors. Each $T_d(p)$ produces a complete $(I, M)$ pair corresponding to disease severity $p$, yielding a large, fully supervised training set that expands smoothly and clinically meaningfully from just a few real labeled samples.

*3) Lattice-of-Experts (LOE) Architecture*

Because pathology progresses at different granularities across diseases and patients, no single $\Delta\gamma$ is universally optimal. To capture these variations in progression scale and structural change, we construct a Lattice-of-Experts (LoE):

$$LoE = \{E_{\Delta\gamma, \alpha}\} \quad (9)$$

Each expert $E_{\Delta\gamma, \alpha}$ is trained under a specific progression granularity $\Delta\gamma$ and interleaving factor $\alpha$. $\Delta\gamma$ sets the sampling resolution along the trajectory, while $\alpha$ controls train–validation alternation during SIV (Sec. III-E). Small $\Delta\gamma$ experts specialize in fine-scale morphological changes; larger $\Delta\gamma$ experts capture global structural transitions. Together, the lattice yields complementary specialists that span the pathological manifold induced by $T_d$.

*4) Self-Reinforcing Interleaved Validation (SIV)*

Few-shot segmentation cannot afford to withhold real anchors for validation, as doing so reduces the effective training signal and risks information leakage. To avoid this, we introduce Self-Reinforcing Interleaved Validation (SIV), which exploit the fully labeled and ordered nature of synthetic set ($V$). For a given expert $E_{\Delta\gamma, \alpha}$, we define a repeating pattern in which training uses $\alpha$ consecutive virtual patients, and the next virtual patient (at the next severity step) is used for validation. Validation samples are thus: 1) Never real anchors (preventing leakage), 2) Never used for training, and 3) Progressively harder; each validation sample is always slightly ahead along pathology trajectory compared to its preceding training segment.

*B. Definition of the Severity Function ($f_d$) and Pathology Anchors (A)*

The generative component of the proposed framework is conditioned on a compact, clinically grounded set of labeled samples that we refer to as the Anchor Set ($A$). The construction of this set is deliberate and principled: anchors are not sampled randomly, but are defined through a disease-specific Severity Function ($f_d$), which quantifies the patient's pathological state.

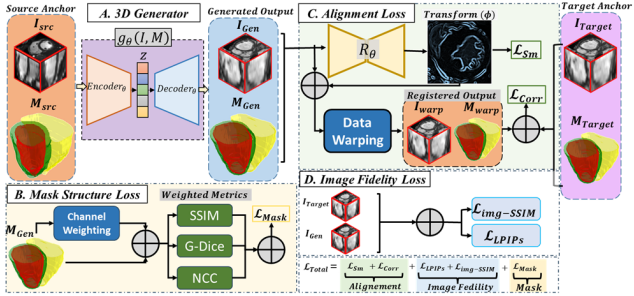

Fig. 2. **Piecewise Latent Space Translation.** Generator $g_\theta$ learns a piecewise latent space between source and target anchors using mask structure, alignment, and image fidelity loss.

### 1) Disease-Specific Severity Function ($f_d$)

We define a severity function $f_d$ that maps each 3D image–mask pair $(I, M)$ to a normalized scalar $\gamma$ (Eq. 3), quantifying severity for disease $d$. Unlike generic similarity metrics, $f_d$ is defined from clinically established biomarkers (e.g., myocardial mass for HCM, ejection fraction for ischemic heart disease), ensuring physiological relevance. We set $\gamma = 0$ at the clinically defined healthy/baseline state and $\gamma = 1$ at the most severe pathological endpoint, estimated from the distribution of clinical measurements (e.g., 5th percentile of healthy, 95th percentile of diseased cohort). Grounding $f_d$ in objective clinical metrics enforces a physiologically ordered progression along $\gamma$. Implementation details of a set of specific per pathology severity functions $f_d$ are provided in Sec. IV.B.1.

### 2) Pathology Anchor Set ($A$)

Using the severity values obtained from $f_d$, we define the Anchor Set as in (Eq.4), which constitutes the entire real labeled dataset used by our method. Unlike conventional few-shot approaches, where the support set contains randomly selected labeled examples, the anchor set is purposefully constructed to include a small number of high-confidence, clinically meaningful points along the disease severity trajectory. Each anchor thus represents an explicitly defined physiological state and serves as a boundary or intermediary constraint for the generative trajectory. The anchor set provides the complete clinical and anatomical supervision required to learn the continuous pathology progression function $T_d$. Implementation details of three different regimes of anchors are in Sec. IV.B.2.

## C. Generative Modeling of the Continuous Latent Pathology Trajectory ($T_d$)

With the pathology anchor set $A$ defined, the next objective is to model the disease progression between anchors through a continuous generative trajectory. As in Eq. 5, we construct a function ($T_d$) that parameterizes the pathological evolution of both image appearance and segmentation anatomy as a function of disease severity. Because only a small number of anchors are available, $T_d$ is learned through a piecewise latent translation framework, in which each segment interpolates between adjacent anchors under explicit clinical constraints as follows:

### 1) Piecewise Latent Space Translation

The anchor set ($A$) defined in (Eq. 4), is composed of anchors $\{a_0, a_1, ..., a_K\}$, each anchor is defined as $a_k = (I_k, M_k, \gamma_k)$. And the anchors are sorted by increasing severity $\gamma_k$. Instead of estimating a global transformation across the entire severity spectrum, we decompose the trajectory into $K$ local translation modules $T_d = \{G_0, G_1, ..., G_{K-1}\}$. Where each generator $G_k$ models nonlinear transformation from anchor $a_k$ to subsequent anchor $a_{k+1}$. Each $G_k$ is implemented as a conditional Generative Adversarial Network (cGAN), trained solely on the two anchors $(a_k, a_{k+1})$. This piecewise modeling stabilizes adversarial training in few-shot regime ensuring that each local transition remains anatomically and clinically plausible.

### 2) 3D Image–Mask Co-Synthesis Architecture

The translation process is performed as shown in Fig. 2, where each anchor $a_k$ comprises an image-mask pair. The trajectory must generate paired image–mask outputs that remain spatially consistent as disease severity changes. To achieve this, we extend the RegGAN [28] to a 3D multi-channel network with deformable refinement. Each module $G_k$ includes:

*A) 3D Generator ($g_{\theta_k}$):* A CycleGAN-inspired encoder–decoder network that jointly synthesizes image–mask pairs from concatenated channels as $(I_{gen}, M_{gen}) = g_{\theta_k}(I_k, M_k)$, enabling joint modeling of appearance and anatomy.

*B) Auxiliary Registration Network ($R_\theta$):* To enforce structural correspondence with anchor $a_{k+1}$. $R_\theta$ predicts a 3D deformation field $\phi_k$, used to warp the generated output as $(I_{warp}, M_{warp}) = (I_{gen} \circ \phi_k, M_{gen} \circ \phi_k)$.

To support pathology-trajectory learning, we adopt: *1) multi-channel* input/output for co-synthesis of 3D images and masks. *2) Perceptual supervision* (replacing conventional GAN [13] discriminator) to stabilize training in low-data regime.

### 3) Training Objectives

Each generator $G_k$ must capture the pathological translation in latent space while simultaneously preserving anatomical integrity. The total loss is defined as:

$$\mathcal{L}_{Total} = \lambda_{Align}\mathcal{L}_{Align} + \lambda_{Fid}\mathcal{L}_{Fid} + \lambda_{Mask}\mathcal{L}_{Mask} \quad (10)$$

A) Alignment Loss ($\mathcal{L}_{Align}$) ensures output agreement with target after deformable refinement, using a smoothness loss $\mathcal{L}_{Sm} = \|\nabla\phi_k\|_2^2$ to regularize deformation field and a correction loss ($\mathcal{L}_{Corr}$) to penalize residual misalignment:

$$\mathcal{L}_{Corr} = \|I_{warp} - I_{k+1}\|_1 + \|M_{warp} - M_{k+1}\|_1 \quad (11)$$

$$\mathcal{L}_{Align} = \lambda_{Corr}\mathcal{L}_{Corr} + \lambda_{Sm}\mathcal{L}_{Sm} \quad (12)$$

B) Image Fidelity Loss: Promotes perceptual realism via deep perceptual similarity [29] and Structural Similarity (SSIM) between generated and target images:

$$\mathcal{L}_{Fid} = \lambda_{LPIPS}\mathcal{L}_{LPIPS} + \mathcal{L}_{img-SSIM} \quad (13)$$

C) Mask Loss: A per-class weighted loss of generalized dice loss, mask SSIM, and Normalized Cross-Correlation, as:

$$\mathcal{L}_{Mask} = \sum_c w_c \left(\mathcal{L}_{GDice}^{(c)} + \mathcal{L}_{ssim}^{(c)} + \mathcal{L}_{NCC}^{(c)}\right) \quad (14)$$

## D. Labeled Virtual Patients Generation (V)

Training the generative trajectory $T_d$ (Eq. 5) yields a continuous, clinically grounded model of disease progression across adjacent anchors. The next objective is to sample the trajectory to construct virtual patient set $V$. While latent-space interpolation, e.g. SLERP [30], produce smooth transitions, it does not ensure uniform data space changes [31], yielding non-uniform severity levels. To obtain evenly spaced, clinically interpretable samples, we introduce a Clinically-Guided Latent Space Sampling strategy (Fig. 3), enforcing almost uniform increments of the severity function $f_d$ (Eq. 4). Main steps are:

*1) Anchors Encoding and Initial Path Generation*

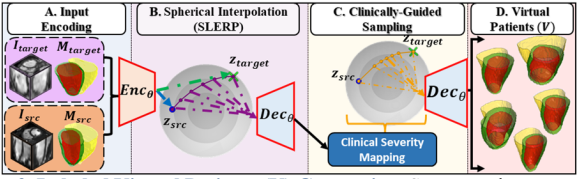

**Fig. 3. Labeled Virtual Patients (V) Generation.** Source and target anchors encoded as latent vectors, and SLERP defines an initial path. Clinical severities are mapped for latent vectors along this path, then, employed to guide resampling, enforcing uniform severity changes. Finally, the resampled latent vectors are decoded to generate labeled 3D virtual patients.

For a trajectory segment $G_k$ relating anchors $a_k$ and $a_{k+1}$, we encode both anchors using the generator's encoder $\text{Enc}_\theta$, as $z_{\text{src}} = \text{Enc}_\theta(I_k, M_k)$ and $z_{\text{tgt}} = \text{Enc}_\theta(I_{k+1}, M_{k+1})$. These latent endpoints define the local region of pathological progression to be sampled. Then, an initial path is formed via spherical linear interpolation (SLERP). For $J$ interpolation weights $\omega_j \in [0,1]$, intermediate vectors are defined as:

$$z_j = \text{SLERP}(z_{\text{src}}, z_{\text{tgt}}; \omega_j), \quad j = 0, \dots, J \quad (15)$$

Each $z_j$ is then decoded into an auxiliary candidate sample, using the generator decoder $\text{Dec}_\theta$ as $(I_j, M_j) = \text{Dec}_\theta(z_j)$.

*2) Clinical Severity Mapping*

To measure the true clinical spacing between SLERP samples, we evaluate the severity of each decoded auxiliary sample using disease-specific severity function $\gamma_j$ defined in Eq.4. Resulting in a discrete mapping between interpolation weights and clinical severities $\{(\omega_j, \gamma_j)\}_{j=0}^{J}$. Because clinical progression is typically nonlinear, $\{\gamma_j\}$ is generally non-uniform, motivating the need for clinically guided resampling.

*3) Clinically-Guided Resampling*

To obtain a clinically uniform set of virtual patients, we define a sequence of $N$ target severity values evenly spaced between the severities of the two anchors denoted as $\gamma_{min}$, $\gamma_{max}$:

$$\Delta\gamma = \frac{\gamma_{\max} - \gamma_{\min}}{N - 1}, \quad \gamma_t^* = \gamma_i + t\,\Delta\gamma \quad (16)$$

Where $t = 0, \dots, N-1$. Next, B-spline interpolation yields new interpolation weights $\{\omega_t^*\}$ matching target severities $\{\gamma_t^*\}$ as: $\{\omega_t^*\} = Bspline(\{\gamma_t^*\})$. The new weights $\{\omega_t^*\}$ define redistributed locations on the initial SLERP path with approximately uniform clinical severity changes ($\gamma_{t+1}^* - \gamma_t^* \approx \Delta\gamma, \forall t$). We define a new clinically calibrated latent vectors set:

$$z_t^* = \text{SLERP}(z_{\text{src}}, z_{\text{tgt}}; \omega_t^*) \quad (17)$$

Finally, decoding yields the fully labeled virtual patient set:

$$V = \{(\text{Dec}_\theta(z_t^*))\}_{t=0}^{N-1} \quad (18)$$

This process produces a dense, clinically uniform sampling of pathological trajectory, generating virtual patient with complete 3D image and segmentation mask, resulting in a large, fully labeled dataset (V) suitable for training the lattice of segmentation experts in a fully supervised manner.

### E. Self-Reinforcing Interleaved Validation (SIV): A Leakage-Free Training Paradigm

A key objective of our framework is to train each segmentation expert using only the few-shot anchor set $A$ and the synthetic, fully-labeled virtual patient set $V$, without withholding any real data for validation. Standard schemes (e.g., k-fold cross-validation) are unsuitable here because (1) they further deplete the already scarce anchor set $A$, and (2) with very few labeled samples, cross-fold reuse introduces information leakage. To address this, we introduce Self-Reinforcing Interleaved Validation (SIV), a leakage-free validation strategy enabled by the properties of $V$: synthetic, fully labeled, and strictly ordered along the clinical severity axis. SIV constructs a dynamic, test-like validation signal directly from the generative trajectory, eliminating the need to reserve real anchors for validation.

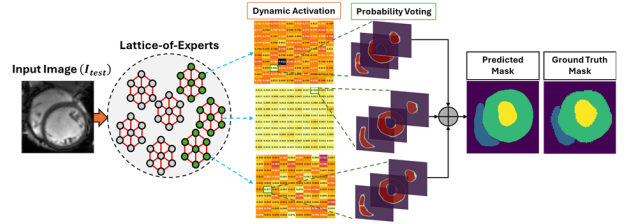

**Fig. 4. Dynamic Activation at inference.** Testing Image fed to trained lattice-of-experts, then, three corresponding scoring maps are generated per label, reflecting specialization of every expert. We employ softmax voting across top performing models per-class to generate final predicted mask.

*1) Interleaved Training–Validation Blocks*

For an expert defined by a granularity–interleaving pair $(\Delta\gamma, \alpha)$, the virtual samples are ordered as $V = \{T_d(k\Delta\gamma)\}_{k=0}^{K_{\max}}$. SIV divides this sequence into repeating blocks consisting of: 1) Training block: the next $\alpha$ consecutive virtual patients 2) Validation sample: the $(\alpha + 1)$-th virtual patient, thus, each expert is trained on increasingly advanced pathological states before being validated on the next, unseen progression step.

*2) Formal Definition of SIV Partitioning*

Given $S_{\text{train}} = A \cup V$, SIV creates the following partitions:

$$S_{\text{train}}^{(\Delta\gamma,\alpha)} = \{T_d(k\Delta\gamma): k \bmod (\alpha + 1) \neq 0\} \cup A \quad (19)$$

$$S_{\text{val}}^{(\Delta\gamma,\alpha)} = \{T_d(k\Delta\gamma): k \bmod (\alpha + 1) = 0\} \quad (20)$$

Key properties are: (1) *Leakage-free:* all validation samples are drawn exclusively from $V$, never from $A$. (2) N*ovel validation samples:* samples used for validation in a given block are never used for training. (3) *Progressively challenging (self-reinforcing):* each validation sample is, by construction, more pathologically advanced than the preceding training samples, providing a consistent out-of-distribution shift. Together, these properties yield a strong, biologically meaningful validation signal without any additional labeled data.

### F. Lattice-of-Experts (LoE) Architecture and Training

*1) Motivation for a Lattice of Specialized Experts*

While SIV enables robust training of a single expert, real-world pathological progression is heterogeneous and disease-dependent: morphological changes range from subtle wall thickening in HCM to large cavity dilation in DCM. A single expert cannot adequately cover this spectrum. Given the continuous generative function $T_d$, there are infinitely many ways to sample the trajectory; training on a narrow sampling strategy leads to over-specialization, whereas training on overly broad samples "averages out" critical structure. To explicitly capture these heterogeneous pathological manifolds, we introduce LoE: a structured, discrete approximation of the continuous generative space.

*2) Construction of the Lattice*

We formulate the generic lattice $\mathcal{L}$ definition (Eq. 9) as:

$$\mathcal{L} = \{E_{\Delta\gamma,\alpha} \mid \Delta\gamma \in \mathcal{D},\ \alpha \in \mathcal{Q}\} \quad (21)$$

where $\mathcal{D}$ is a set of severity progression granularities and $\mathcal{Q}$ a set of SIV interleaving factors. Each pair $(\Delta\gamma, \alpha)$ specifies a unique training schedule and validation cadence, and each expert is trained independently from scratch on the fully labeled set $S_{\text{train}}$ using its SIV split. This structured diversity promotes specialization: small $\Delta\gamma$ experts capture fine structural changes (early/borderline pathology), whereas large $\Delta\gamma$ experts model broad remodeling. Likewise, small $\alpha$ yields frequent validation and reduced overfitting, while large $\alpha$ supports generalization over long pathological transitions. The resulting lattice is a clinically indexed manifold, increasing the likelihood that an "appropriate" expert exists for an OOD test image.

*G. Test-Time Dynamic Activation of Lattice-of-Experts*

At inference, PathCo-LatticE acts as a zero-shot, self-configuring system: it requires no test-time fine-tuning, adaptation, or additional labeled/unlabeled data. A test image $I_{\text{test}}$ is passed in parallel through all experts $E_m \in \mathcal{L}$, each producing a multiclass probability map. Rather than selecting a single global expert, we perform per-class selection (RV, myocardium, LV cavity), allowing the most specialized expert to be chosen for each structure. To pick the best expert for class $c$, we define a proxy score based solely on softmax probabilities: an expert whose segmentation best matches the anatomy should yield confident, high probabilities within its own foreground mask. Let $y_E = E(I_{\text{test}})$ denote the full multiclass probability tensor, $y_{E,c}$ the probability map for class $c$, and $y'_{E,c}$ the corresponding binarized mask defined as:

$$y'_{E,c}(v) = \mathbf{1}\left\{\arg\max_k y_{E,k}(v) = c\right\} \quad (22)$$

We define a proxy score for expert $E$ and class $c$ as the mean foreground probability of class $c$ over its own predicted region:

$$\Psi(I_{\text{test}}, E, c) = \frac{1}{|y'_{E,c}|} \sum_{v \in y'_{E,c}} y_{E,c}(v) \quad (23)$$

This score increases when the predicted foreground is internally consistent, probabilities are sharp and confident, and the expert captures class-specific morphology it was trained to specialize in. Thus, $\Psi$ functions as a supervised-free surrogate to identify which expert is most aligned with the anatomy in the test image. For each class $c$, the optimal expert is:

$$E_c^* = \arg\max_{E \in \mathcal{L}} \Psi(I_{\text{test}}, E, c) \quad (24)$$

In practice, this yields three experts per test image—$E_{\text{RV}}^*, E_{\text{Myo}}^*$, and $E_{\text{Pool}}^*$—each selected independently from the lattice. The final segmentation $y_{\text{final}}$ is obtained by fusing only probability maps of these three activated experts via voxel wise maximum-probability selection. This dynamic per-image, per-chamber activation (Fig. 4) enables each test case to draw upon the expert most aligned with its implied pathological severity and structural characteristics, enabling robust generalization across vendors, protocols, and unseen domains.

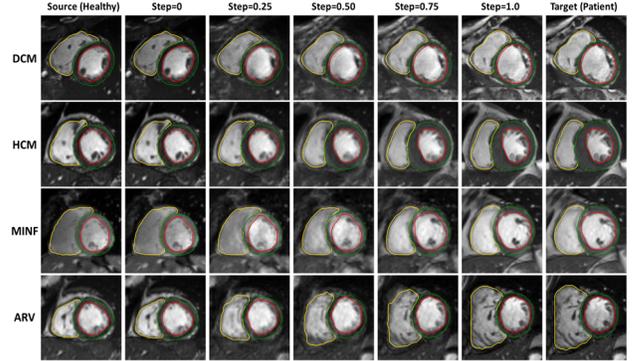

**Fig. 5. Examples of Virtual Patients.** PathCo-LatticE synthetic virtual patients generated from trajectories: DCM, HCM, MINF, and ARV, from a healthy (left) to a patient (right) anchor, showing LV pool (red), Myo (green), and RV (yellow), illustrating smooth, clinically plausible severity transitions with high image–mask correspondence.

## IV. EXPERIMENTS AND RESULTS

*A. Datasets and Evaluation Metrics*

We evaluated PathCo-LatticE performance and generalizability using two distinct datasets in a single-source domain generalization setting:

*In-Distribution Training Dataset:* ACDC dataset [5] includes CINE short-axis (SAX) MRI scans for 20 healthy subjects and 80 patients, including four pathologies: dilated cardiomyopathy (DCM), hypertrophic cardiomyopathy (HCM), myocardial infarction (MINF), and abnormal right ventricle (ARV). Manual contours provided at end-diastolic (ED) and end-systolic (ES) for LV blood pool (Pool), myocardium (Myo), and right ventricle (RV). Following common practice [8], ACDC split into train:validation:test with ratio 7:1:2.

*Out-of-Distribution Evaluation Dataset:* To evaluate generalizability, we used M&Ms dataset [4], that includes CINE SAX MRI scans from six centers and four scanner vendors (Siemens, Philips, GE, Canon). M&Ms includes healthy group and broad spectrum of pathologies including DCM, HCM, ARV, and additional pathologies such as hypertensive heart disease (HHD), athlete's heart syndrome (AHS), left ventricular non-compaction (LVNC). Of 345 subjects, we excluded 25 unannotated, yielding 320 subjects with manual ED/ES contours. None of M&Ms cases were used for training/validation; M&Ms served as an out-of-distribution test set. Segmentation performance is measured using 3D Dice (%) and HD95 (in mm) per class.

*B. Implementation Details*

*1) Clinical Instantiation of Severity Functions ($f_d$)*

The abstract severity function $f_d$ (Eq. 3) was instantiated for ACDC pathologies (DCM, HCM, MINF, ARV) using clinically validated biomarkers computed from segmentation masks $M \in \{M_{LV}, M_{Myo}, M_{RV}\}$. For each pathology, we modeled a dedicated latent trajectory driven by canonical imaging surrogates of disease progression. All functional forms and normalization statistics were computed *only* on ACDC training split preserving zero-shot validity on M&Ms. Let $V(\cdot)$ denote volume and $L(\cdot)$ axis length. We define severity functions ($f$):

*DCM Severity function ($f_{DCM}$):* To capture the trajectory of DCM progression, we used LV sphericity index at end-diastole, since DCM is characterized by progressive LV dilatation and geometric remodeling toward a more spherical LV [2]. Thus, increasing short-to-long axis ratios reflect worsening severity:

$$f_{DCM}(M) = \frac{L_{short}(M_{LV}^{ED})}{L_{long}(M_{LV}^{ED})} \quad (25)$$

*HCM Severity function ($f_{HCM}$):* To capture the trajectory of HCM progression, we used LV myocardial mass (volume) at end-diastole, as HCM severity is dominated by hypertrophic wall thickening and increased myocardial burden [1]:

$$f_{HCM}(M) = V(M_{Myo}^{ED}) \quad (26)$$

*MINF Severity function ($f_{MINF}$):* To capture the trajectory of MINF progression, we used LV ejection fraction computed from ED/ES LV cavity volumes, because infarction-related remodeling primarily manifests as progressive systolic dysfunction and reduced pump efficiency [32]:

$$f_{MINF}(M) = \frac{V(M_{LV}^{ED}) - V(M_{LV}^{ES})}{V(M_{LV}^{ED})} \quad (27)$$

*ARV Severity function ($f_{ARV}$):* To capture the trajectory of ARV progression, we used RV cavity volume at end-diastole, reflecting the defining phenotype of progressive RV dilatation/volume overload in this cohort [3]:

$$f_{ARV}(M) = V(M_{RV}^{ED}) \quad (28)$$

Although we instantiate anchors from the four ACDC pathologies, PathCo-LatticE is not limited to these classes: the trajectories define generic remodeling axes (dilatation, hypertrophy, systolic dysfunction, RV enlargement) along which many cardiac diseases lie, singly or in combination. By encoding these biomarker-driven progressions, the model learns a latent disease hyperspace that extends beyond ACDC; we demonstrate this under a strict zero-shot protocol on M&Ms, segmenting pathologies absent from ACDC while keeping all anchors and severity statistics derived solely from the ACDC , results are discussed in (Sec. IV.C-6).

### 2) Pathology Anchor Set (A)

To define the labeled anchor set in Eq. 4, we first specify the severity scale $\gamma$ in Eq. 3. Heterogeneous, often non-Gaussian biomarkers are mapped to $\gamma \in [0,1]$ using a percentile-based scheme, where the $k$-th percentile $P_k$ corresponds to $\gamma \approx k/100$. This non-parametric mapping is robust to skewed clinical distributions (e.g., myocardial volume), where mean-based standardization is unstable. We further set trajectory endpoints at $P_5$ and $P_{95}$ rather than absolute minima/maxima to reduce the influence of outliers.

Anchor selection is performed independently for each pathology $p \in \mathcal{P} = \{DCM, HCM, MINF, ARV\}$. For each $p$, we define a healthy baseline anchor $a_p^{P_5}$ as the healthy subject closest to $P_5 (\gamma \approx 0.05)$, and a severe endpoint anchor $a_p^{P_{95}}$ as the pathology subject at $P_{95}(\gamma \approx 0.95)$. To assess scalability, we then consider three cumulative anchor-labeling regimes with progressively finer trajectory resolution.

*i) $A_7$ (Coarse/endpoint regime):* anchors at $\gamma \in \{0.05, 0.95\}$.

$$A_7 = \bigcup_{p \in \mathcal{P}} \{a_p^{P_5}, a_p^{P_{95}}\} \quad (29)$$

Yielding 8 anchors across four pathologies in principle (4 pathologies × 2 anchors), but one healthy subject from ACDC coincided with $P_5$ for two severity functions, resulting in a total of 7 unique labeled anchors.

*Table. I In-Distribution (ACDC) Results Comparison with SOTA FSL Methods*

| Method | Number of Subjects (ACDC) | | Dice ↑ | | | | HD95 (mm) ↓ | | | |
|---|---|---|---|---|---|---|---|---|---|---|
| | Labeled | Unlabeled | Pool | Myo | RV | Avg | Pool | Myo | RV | Avg |
| SS-Net [21] | | 63(90%) | 0.891 | 0.763 | 0.756 | 0.803 | 5.73 | 8.45 | 10.27 | 8.15 |
| MC-Net [23] | | 63(90%) | 0.893 | 0.796 | 0.765 | 0.818 | 5.20 | 6.72 | 10.32 | 7.41 |
| BCP [9] | 7(10%) | 63(90%) | 0.933 | 0.817 | 0.847 | 0.866 | 3.59 | 5.62 | 9.93 | 6.38 |
| PPC [8] | | 63(90%) | 0.920 | 0.820 | 0.879 | 0.873 | 3.40 | 4.02 | 6.97 | 4.80 |
| Ours | | - | 0.935 | 0.864 | 0.906 | 0.902 | 2.50 | 2.37 | 7.96 | 4.27 |
| SS-Net [21] | | 59(84%) | 0.895 | 0.799 | 0.790 | 0.828 | 5.36 | 6.21 | 8.40 | 6.66 |
| MC-Net [23] | | 59(84%) | 0.909 | 0.807 | 0.804 | 0.840 | 5.07 | 6.26 | 7.98 | 6.44 |
| BCP [9] | 11(16%) | 59(84%) | 0.929 | 0.809 | 0.859 | 0.866 | 4.30 | 6.02 | 9.89 | 6.74 |
| PPC [8] | | 59(84%) | 0.934 | 0.845 | 0.883 | 0.887 | 3.05 | 3.60 | 7.69 | 4.78 |
| Ours | | - | 0.934 | 0.871 | 0.917 | 0.907 | 2.46 | 2.19 | 6.75 | 3.80 |
| SS-Net [21] | | 51(73%) | 0.915 | 0.811 | 0.806 | 0.844 | 4.14 | 5.72 | 7.55 | 5.80 |
| MC-Net [23] | | 51(73%) | 0.932 | 0.816 | 0.839 | 0.862 | 3.04 | 5.56 | 6.35 | 4.98 |
| BCP [9] | 19(27%) | 51(73%) | 0.940 | 0.832 | 0.873 | 0.882 | 3.65 | 5.00 | 9.11 | 5.92 |
| PPC [8] | | 51(73%) | 0.939 | 0.838 | 0.870 | 0.882 | 3.26 | 3.76 | 7.75 | 4.92 |
| Ours | | - | 0.938 | 0.875 | 0.926 | 0.913 | 2.50 | 2.34 | 6.11 | 3.65 |

*Table. II Zero-Shot Out-of-Distribution Generalizability (M&Ms) Comparison with SOTA FSL Methods*

| Method | Number of Subjects (ACDC) | | Dice ↑ | | | | HD95 (mm) ↓ | | | |
|---|---|---|---|---|---|---|---|---|---|---|
| | Labeled | Unlabeled | Pool | Myo | RV | Avg | Pool | Myo | RV | Avg |
| SS-Net [21] | | 63(90%) | 0.866 | 0.726 | 0.728 | 0.773 | 5.553 | 6.886 | 11.428 | 7.956 |
| MC-Net [23] | | 63(90%) | 0.888 | 0.770 | 0.731 | 0.796 | 4.611 | 5.106 | 11.454 | 7.057 |
| BCP [9] | 7(10%) | 63(90%) | 0.903 | 0.761 | 0.775 | 0.813 | 3.598 | 5.621 | 9.930 | 6.383 |
| PPC [8] | | 63(90%) | 0.903 | 0.788 | 0.831 | 0.841 | 3.402 | 4.029 | 6.978 | 4.803 |
| Ours | | - | 0.927 | 0.851 | 0.871 | 0.883 | 2.505 | 2.372 | 7.961 | 4.279 |
| SS-Net [21] | | 59(84%) | 0.867 | 0.762 | 0.742 | 0.790 | 6.166 | 5.275 | 11.402 | 7.614 |
| MC-Net [23] | | 59(84%) | 0.891 | 0.793 | 0.760 | 0.815 | 4.446 | 4.271 | 10.113 | 6.276 |
| BCP [9] | 11(16%) | 59(84%) | 0.890 | 0.749 | 0.793 | 0.811 | 4.306 | 6.028 | 9.893 | 6.742 |
| PPC [8] | | 59(84%) | 0.912 | 0.808 | 0.831 | 0.850 | 3.057 | 3.604 | 7.691 | 4.784 |
| Ours | | - | 0.925 | 0.857 | 0.887 | 0.890 | 2.469 | 2.191 | 6.750 | 3.803 |
| SS-Net [21] | | 51(73%) | 0.883 | 0.781 | 0.755 | 0.806 | 5.032 | 4.682 | 10.345 | 6.686 |
| MC-Net [23] | | 51(73%) | 0.897 | 0.783 | 0.796 | 0.825 | 4.003 | 4.480 | 8.706 | 5.730 |
| BCP [9] | 19(27%) | 51(73%) | 0.903 | 0.771 | 0.796 | 0.823 | 3.655 | 5.005 | 9.112 | 5.924 |
| PPC [8] | | 51(73%) | 0.910 | 0.801 | 0.818 | 0.843 | 3.263 | 3.764 | 7.759 | 4.928 |
| Ours | | - | 0.925 | 0.861 | 0.899 | 0.895 | 2.509 | 2.347 | 6.114 | 3.656 |

*ii) $A_{11}$ (Medium/median regime):* adds median state $\gamma \approx 0.50$ for each pathology:

$$A_{11} = A_7 \cup \bigcup_{p \in \mathcal{P}} \{a_p^{P_{50}}\} \tag{30}$$

Adding 4 subjects across the 4 pathologies (4 pathologies × 1 anchors), resulting in a total of 11 unique labeled anchors.

*iii) $A_{19}$ (High/quartile regime):* refines trajectory by adding $P_{25}$ and $P_{75}$ anchors for each pathology:

$$A_{19} = A_{11} \cup \bigcup_{p \in \mathcal{P}} \{a_p^{P_{25}}, a_p^{P_{75}}\} \tag{31}$$

Adding 8 subjects across the 4 pathologies (4 pathologies × 2 anchors) resulting in a total of 19 unique labeled anchors.

### 3) Virtual Patients (V) 3D Image-Mask Co-Synthesis

We model each piecewise latent space as a modified multi-channel 3D RegGAN, training one generator per anchor pair and cardiac phase (ED/ES), the four input channels: Image, Pool, Myo and RV. All anchor pairs were rigidly registered, resampled to 1.5×1.5 mm, cropped with a 20% margin, resized to 256×256×16, and normalized to [0,1]. RegGAN models were trained for 1000 epochs with Adam (LR=2×10⁻⁵) using weighted loss (Eq.10): $\lambda_{Align}$, $\lambda_{Fid} = 1$, $\lambda_{Mask} = 20$, alignment loss (Eq.12): $\lambda_{Corr} = 20$, $\lambda_{Sm} = 40$, image fidelity loss (Eq.13): $\lambda_{LPIPS} = 1$, $\lambda_{img-SSIM} = 10$. Mask loss (Eq.14) uses per-channel weights $w_c$ as the inverse of relative volume, normalized between [0,1], to balance channel weights. We generate virtual patients by encoding anchors with RegGAN encoder, applying SLERP ($J = 200$, Eq.15), and resampling using clinical functions (Eqs.25-28). Across anchoring regimes ($A_7, A_{11}, A_{19}$), this yields 400, 1200, and 2000 3D virtual patients (V) (≈12.8K, 38.4K, 64K 2D images). Fig. 5 Shows generated virtual patients examples.

### 4) Lattice-of-Experts Training and Test-Time Activation

We construct the LoE (Eq. 21) using 10 granularity levels $\mathcal{D} = \{0.1, 0.2, ..., 1.0\}$ and reinforcement $\alpha \in Q$ steps, s.t. $Q = \{2,4,6,8,10\}$, yielding 100 nnU-Net [33] models, each trained for 100 epochs on a unique synthetic configuration. At inference, each image passes through all experts, producing per-label probability maps; top expert is selected via a scoring function (Eq.23, 24), final predictions are obtained by softmax voting over top 3 models. Experiments ran on Nvidia H100 GPU with CUDA 12.4 and Pytorch 2.5.1.

## C. Segmentation & Generalizability Evaluation

### 1) Experimental Setup and Baselines Comparison

We benchmarked PathCo-LatticE against four SOTA few-shot segmentation methods: SS-Net[20], MC-Net[22], BCP[8], and PPC[7]. All methods were trained using identical ACDC splits (70 training, 10 validation subjects) and the same preprocessing pipeline (including center cropping) to ensure a strict one-to-one comparison. We evaluated three labeling regimes using 7 (10%), 11 (16%), and 19 (27%) labeled ACDC subjects. Critically, baseline methods additionally exploited remaining ACDC training subjects as unlabeled data to align source distributions. In contrast, PathCo-LatticE used no unlabeled data, relying exclusively on the limited labeled anchors and the virtual patients cohort for supervision.

### 2) In-Distribution Performance (ACDC)

Table I summarizes performance on the ACDC test set. Across all labeling regimes, PathCo-LatticE consistently outperformed all FSL baselines. In the most challenging low-data regime (7 labeled subjects), our method surpassed the strongest baseline (PPC) by 2.9% Dice and reduced HD95 by 11%. Furthermore, baseline methods tend to saturate or even degrade when the number of labeled subjects increases from 11 to 19, suggesting that their pseudo-labeling mechanisms reach a noise floor. In contrast, PathCo-LatticE exhibits a monotonic performance increase with additional labeled anchors, indicating that extra anchors refine the latent trajectories and yield more informative synthetic supervision.

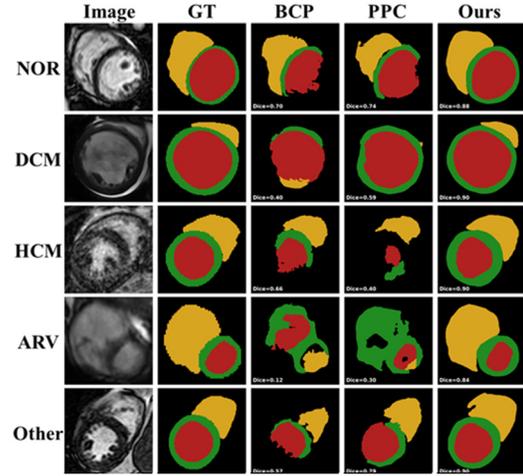

Fig. 6. Qualitative Results for Zero-Shot Out-of-Distribution (M&Ms). PathCo-LatticE predictions versus two FSL methods (BCP [8], PPC [7]) and ground truth, with Pool (red), Myo (green), and RV (yellow). Our method shows better generalization, improving Myo and RV contours over SOTA FSL.

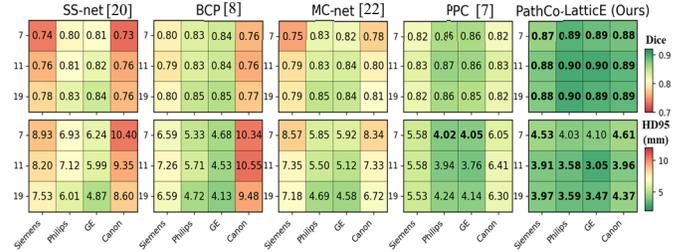

Fig. 7. Zero-shot Vendor Harmonization (M&M). PathCo-LatticE shows higher, harmonized performance across all vendors compared to SOTA FSL methods, on all labeled-data settings for average Dice (top) and HD95 (bottom).

### 3) Zero-Shot Out-of-Distribution Generalization

We next evaluated zero-shot generalization by training on single-center single-vendor ACDC only and testing directly on the OOD multi-center multi-vendor M&Ms dataset (N=320) without target-domain fine-tuning (Table II). PathCo-LatticE achieved the highest out-of-distribution robustness, with an average dice improvement of 5.2 and an HD95 reduction of 25.8% compared with the strongest FSL baseline (Table II). Conventional semi-supervised FSL methods assume that the unlabeled pool is drawn from a distribution similar to the labeled training data; under domain shift, this assumption breaks down, causing performance deterioration. In contrast,

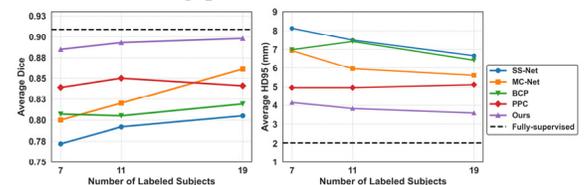

Fig. 8. Comparison to full-supervised training (M&M). PathCo-LatticE approaches fully-supervised (trained on 100% M&M), outperforming SOTA FSL methods, with increasing performance as more labeled subjects are added.

PathCo-LatticE generates a virtual cohort constrained by physiological biomarkers rather than raw image statistics. As shown in Fig. 6, the network was able to learn anatomical changes (e.g., geometric deviations from normal sphericity) instead of overfitting to scanner-specific texture, resulting in improved stability across unseen domains.

*4) Zero-shot Vendor Harmonization*

To analyze robustness to acquisition variability, we stratified zero-shot M&Ms performance by the four scanner vendor (Siemens, Philips, GE, Canon). Heatmaps in Fig. 7 show that while baselines FSL methods exhibit substantial vendor-to-vendor variability and marked degradation on unseen vendors (e.g., Canon), PathCo- LatticE maintains uniform and highest performance across all four vendors. This supports PathCo-LatticE's pathology-constrained generator in implicitly augmenting the virtual patients with diverse "styles" while keeping anatomy tied to the severity function, jointly modeling structure and style to reduce vendor-induced domain shifts enabling harmonization without target-domain adaptation.

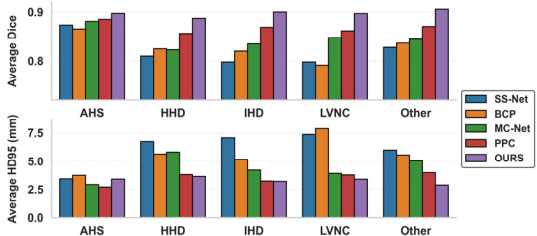

Fig. 9. **Zero-shot generalizability to unseen pathologies.** Our framework achieves the highest Average Dice (top) and HD95 (bottom) across unseen pathologies (AHS, HHD, IHD, LVNC, and Other), consistently outperforming competing approaches. This demonstrates strong zero-shot generalization to entirely unseen cardiac disease phenotypes.

*5) Bridging the Gap to Full Supervision*

To contextualize the utility of the proposed framework, we compared PathCo-LatticE (trained on 19 labeled ACDC subjects) to a fully supervised upper bound trained on 100% of the M&Ms training set (N=184). As in Fig. 8, the zero-shot PathCo-LatticE model narrows the gap to within 1% Dice and 1.5 mm HD95 relative to the fully supervised model. Baseline FSL methods lag 5–11% behind and exhibit substantially larger boundary errors. These results indicate that the pathology-constrained synthetic cohort effectively compensates for the limited number of labeled anchors. While not eliminating the value of target-domain labels, this suggests that pathology-aware synthesis can substantially reduce the need for extensive manual annotation in the target domain.

*6) Generalization to Unseen Pathologies*

We further evaluated and compared generalization to M&Ms five sub-cohorts of pathologies not present in ACDC training set (e.g. Ischemic Disease, LV Non-Compaction). Fig. 9, shows that PathCo-LatticE ranks first across all five unseen pathology groups, with notable HD95 reductions for heterogeneous "Other" (28.1%) and LVNC (10.4%) categories relative to the strongest FSL baseline. These findings support our design of the "Latent Hyperspace," where anchors defined for four primary pathologies (DCM, HCM, MINF, ARV) instantiate principal remodeling axes such as dilation, hypertrophy, and systolic dysfunction. By encoding biomarker-driven trajectories along these axes, the model learns a latent space that can represent combinatorial phenotypes. LVNC, for example, shares geometric features with both hypertrophy (wall thickening) and dilation, explaining the improved segmentation performance even without LVNC labels during training.

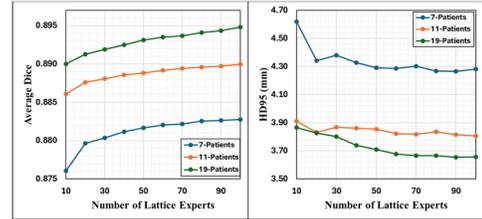

Fig. 10. **Ablation for number of experts in the Lattice** Showing the effectiveness of using more experts to achieve better performance on the zero-shot Out-of-Distribution dataset (M&M; N=320).

*7) Ablation Study*

We evaluated the impact of the Lattice-of-Experts (LoE) and increasing number of nodes in the lattice (Fig. 10). Increasing expert density—finer sampling of latent disease sub-regions—systematically improves OOD Dice and HD95 (Fig. 10). Notably even at 10 nodes, our performance exceeded that of FSL baselines (Table II). A single monolithic network must trade off sensitivity to early subtle changes versus robustness to late-stage remodeling. LoE allocates small-granularity experts to fine textures/early remodeling and coarse-granularity experts to pronounced shape changes, while dynamic test-time activation selects the most physiologically aligned expert(s), improving segmentation across diverse pathologies.

## V. Discussion and Conclusions

The superior performance of PathCo-LatticE, particularly in the low-data regime, highlights a critical advantage over current semi-supervised FSL approaches. While baseline FSL methods exhibited performance saturation—likely due to pseudo-labeling noise—our framework maintained a monotonic upward trajectory. This suggests that our pathology-constrained synthesis effectively bridges the "spectral gaps" in sparse datasets. By populating these gaps with continuous disease trajectories, the network encounters a broader range of anatomical deformations, enhancing generalizability.

Crucially, the PathCo-LatticE framework achieved superior uniform and harmonized zero-shot performance across unseen vendors without target-domain adaptation on the challenging multi-center and multi-vendor OOD M&Ms dataset. Unlike conventional FSL, which fails when the unlabeled pool drifts from the training distribution, our approach grounds learning in physiological biomarkers. This encourages the model to learn anatomical invariants (e.g., geometric remodeling) rather than overfitting to vendor-specific artifacts.

This robustness extends to unseen pathologies, such as LV Non-Compaction (LVNC), supporting the "latent hyperspace" hypothesis. By encoding fundamental remodeling axes (e.g., dilatation, hypertrophy), PathCo-LatticE learns a latent space capable of representing combinatorial phenotypes. Thus, complex conditions like LVNC are positioned appropriately within this space despite their absence from the training set.

Beyond synthesis, two architectural innovations proved vital. First, Self-Reinforcing Interleaved Validation (SIV) provides a leakage-free validation paradigm for small-anchor regimes. By utilizing ordered virtual samples, SIV enables principled model selection without withholding scarce real anchors. Second, the Lattice-of-Experts (LoE) decomposes the task into specialized sub-regions of the disease spectrum. Ablation studies confirm that increasing expert density systematically improves OOD performance by reconciling sensitivity to early remodeling with robustness to advanced pathology. In our experiments, this strategy narrowed the gap to fully supervised upper bounds to within ~1% Dice and 1.5 mm HD95, suggesting that pathology-constrained synthesis acts as a practical, scalable surrogate for large-scale annotation. Notably, PathCo-LatticE severity-based formulation is mathematically generic and adaptable to any domain with quantifiable priors (e.g., tumor size in oncology or atrophy in neuroimaging).

In conclusion, PathCo-LatticE reformulates few-shot segmentation into a fully supervised paradigm using pathology-constrained synthesis. By combining continuous virtual cohorts, Self-Reinforcing Interleaved Validation, and a dynamic Lattice-of-Experts, it achieves state-of-the-art zero-shot generalization on the multi-vendor M&Ms dataset. The framework harmonizes performance across unseen domains and narrows the gap to fully supervised models to within ~1% Dice, supporting physiologically constrained synthesis as a scalable alternative to massive annotation.

REFERENCES

[1]  B. J. Maron, S. R. Ommen, C. Semsarian, P. Spirito, I. Olivotto, and M. S. Maron, "Hypertrophic Cardiomyopathy," *J. Am. Coll. Cardiol.*, vol. 64, no. 1, pp. 83–99, Jul. 2014.
[2]  Y. Liang *et al.*, "Left Ventricular Spherical Index Is an Independent Predictor for Clinical Outcomes in Patients With Nonischemic Dilated Cardiomyopathy," *JACC Cardiovasc. Imaging*, vol. 12, no. 8, pp. 1578–1580, Aug. 2019.
[3]  J. Sanz, D. Sánchez-Quintana, E. Bossone, H. J. Bogaard, and R. Naeije, "Anatomy, Function, and Dysfunction of the Right Ventricle," *J. Am. Coll. Cardiol.*, vol. 73, no. 12, pp. 1463–1482, Apr. 2019.
[4]  V. M. Campello *et al.*, "Multi-Centre, Multi-Vendor and Multi-Disease Cardiac Segmentation: The M&Ms Challenge," *IEEE Trans. Med. Imaging*, vol. 40, no. 12, pp. 3543–3554, Dec. 2021.
[5]  O. Bernard *et al.*, "Deep learning techniques for automatic MRI cardiac multi-structures segmentation and diagnosis: is the problem solved?," *IEEE Trans. Med. Imaging*, vol. 37, no. 11, pp. 2514–2525, 2018.
[6]  L. Zhang *et al.*, "Generalizing deep learning for medical image segmentation to unseen domains via deep stacked transformation," *IEEE Trans. Med. Imaging*, vol. 39, no. 7, pp. 2531–2540, 2020.
[7]  Y. Yuan, X. Wang, X. Yang, and P.-A. Heng, "Effective Semi-Supervised Medical Image Segmentation With Probabilistic Representations and Prototype Learning," *IEEE Trans. Med. Imaging*, 2024.
[8]  Y. Bai, D. Chen, Q. Li, W. Shen, and Y. Wang, "Bidirectional copy-paste for semi-supervised medical image segmentation," in *Proceedings of the IEEE/CVF conference on computer vision and pattern recognition*, 2023, pp. 11514–11524.
[9]  H. Jiang, M. Gao, H. Li, R. Jin, H. Miao, and J. Liu, "Multi-learner based deep meta-learning for few-shot medical image classification," *IEEE J. Biomed. Health Inform.*, vol. 27, no. 1, pp. 17–28, 2022.
[10] Y. Feng, Y. Wang, H. Li, M. Qu, and J. Yang, "Learning what and where to segment: A new perspective on medical image few-shot segmentation," *Med. Image Anal.*, vol. 87, p. 102834, Jul. 2023.
[11] E. Pachetti and S. Colantonio, "A systematic review of few-shot learning in medical imaging," *Artif. Intell. Med.*, vol. 156, p. 102949, Oct. 2024.
[12] D. P. Kingma and M. Welling, "An introduction to variational autoencoders," *Found. Trends® Mach. Learn.*, vol. 12, no. 4, pp. 307–392, 2019.
[13] I. Goodfellow *et al.*, "Generative adversarial networks," *Commun. ACM*, vol. 63, no. 11, pp. 139–144, Oct. 2020.
[14] Z. Zhang, M. Li, and J. Yu, "On the convergence and mode collapse of GAN," in *SIGGRAPH Asia 2018 Technical Briefs*, Tokyo Japan: ACM, Dec. 2018, pp. 1–4.
[15] Y. Chen *et al.*, "Generative adversarial networks in medical image augmentation: a review," *Comput. Biol. Med.*, vol. 144, p. 105382, 2022.
[16] J.-Y. Zhu, T. Park, P. Isola, and A. A. Efros, "Unpaired image-to-image translation using cycle-consistent adversarial networks," in *Proceedings of the IEEE international conference on computer vision*, 2017, pp. 2223–2232.
[17] "XCAT-GAN for Synthesizing 3D Consistent Labeled Cardiac MR Images on Anatomically Variable XCAT Phantoms," in *Lecture Notes in Computer Science*, Cham: Springer International Publishing, 2020, pp. 128–137.
[18] N. Konz, Y. Chen, H. Dong, and M. A. Mazurowski, "Anatomically-Controllable Medical Image Generation with Segmentation-Guided Diffusion Models," in *Medical Image Computing and Computer Assisted Intervention – MICCAI 2024*, vol. 15007, M. G. Linguraru, Q. Dou, A. Feragen, S. Giannarou, B. Glocker, K. Lekadir, and J. A. Schnabel, Eds., in Lecture Notes in Computer Science, vol. 15007. , Cham: Springer Nature Switzerland, 2024, pp. 88–98.
[19] J. Snell, K. Swersky, and R. Zemel, "Prototypical networks for few-shot learning," *Adv. Neural Inf. Process. Syst.*, vol. 30, 2017.
[20] Y. Wu, Z. Wu, Q. Wu, Z. Ge, and J. Cai, "Exploring Smoothness and Class-Separation for Semi-supervised Medical Image Segmentation," in *Medical Image Computing and Computer Assisted Intervention – MICCAI 2022*, vol. 13435, L. Wang, Q. Dou, P. T. Fletcher, S. Speidel, and S. Li, Eds., in Lecture Notes in Computer Science, vol. 13435. , Cham: Springer Nature Switzerland, 2022, pp. 34–43.
[21] W. Zhao *et al.*, "MetaSSL: A General Heterogeneous Loss for Semi-Supervised Medical Image Segmentation," *IEEE Trans. Med. Imaging*, 2025.
[22] Y. Wu, M. Xu, Z. Ge, J. Cai, and L. Zhang, "Semi-supervised Left Atrium Segmentation with Mutual Consistency Training," in *Medical Image Computing and Computer Assisted Intervention – MICCAI 2021*, vol. 12902, M. De Bruijne, P. C. Cattin, S. Cotin, N. Padoy, S. Speidel, Y. Zheng, and C. Essert, Eds., in Lecture Notes in Computer Science, vol. 12902. , Cham: Springer International Publishing, 2021, pp. 297–306.
[23] Y. Wang, L. Zhang, Y. Yao, and Y. Fu, "How to trust unlabeled data? instance credibility inference for few-shot learning," *IEEE Trans. Pattern Anal. Mach. Intell.*, vol. 44, no. 10, pp. 6240–6253, 2021.
[24] I. Tougui, A. Jilbab, and J. El Mhamdi, "Impact of the choice of cross-validation techniques on the results of machine learning-based diagnostic applications," *Healthc. Inform. Res.*, vol. 27, no. 3, pp. 189–199, 2021.
[25] R. A. Jacobs, M. I. Jordan, S. J. Nowlan, and G. E. Hinton, "Adaptive mixtures of local experts," *Neural Comput.*, vol. 3, no. 1, pp. 79–87, 1991.
[26] Y. Sun, X. Wang, Z. Liu, J. Miller, A. Efros, and M. Hardt, "Test-time training with self-supervision for generalization under distribution shifts," in *International conference on machine learning*, PMLR, 2020, pp. 9229–9248.
[27] D. Wang, E. Shelhamer, S. Liu, B. Olshausen, and T. Darrell, "Tent: Fully Test-Time Adaptation by Entropy Minimization," in *International Conference on Learning Representations*.
[28] L. Kong, C. Lian, D. Huang, Y. Hu, and Q. Zhou, "Breaking the dilemma of medical image-to-image translation," *Adv. Neural Inf. Process. Syst.*, vol. 34, pp. 1964–1978, 2021.
[29] R. Zhang, P. Isola, A. A. Efros, E. Shechtman, and O. Wang, "The Unreasonable Effectiveness of Deep Features as a Perceptual Metric," in *2018 IEEE/CVF Conference on Computer Vision and Pattern Recognition*, Salt Lake City, UT: IEEE, Jun. 2018, pp. 586–595.
[30] T. White, "Sampling Generative Networks," Dec. 06, 2016, *arXiv*: arXiv:1609.04468.
[31] G. Arvanitidis, L. K. Hansen, and S. Hauberg, "Latent space oddity: On the curvature of deep generative models," in *6th International Conference on Learning Representations, ICLR 2018*, 2018.
[32] J. A. Panza *et al.*, "Myocardial Viability and Long-Term Outcomes in Ischemic Cardiomyopathy," *N. Engl. J. Med.*, vol. 381, no. 8, pp. 739–748, Aug. 2019.
[33] F. Isensee, P. F. Jaeger, S. A. Kohl, J. Petersen, and K. H. Maier-Hein, "nnU-Net: a self-configuring method for deep learning-based biomedical image segmentation," *Nat. Methods*, vol. 18, no. 2, pp. 203–211, 2021.